 \documentclass[final,5p,times,twocolumn]{elsarticle}

\usepackage{graphicx}
\usepackage{hyperref}
\usepackage{float}

\usepackage{amssymb}
\makeatletter
\def\ps@pprintTitle{%
  \let\@oddhead\@empty
  \let\@evenhead\@empty
  \def\@oddfoot{\reset@font\hfil\thepage\hfil}
  \let\@evenfoot\@oddfoot
}
\makeatother

\begin{document}

\begin{frontmatter}



%%\title{Test}

\title{\vspace{-15mm}\fontsize{24pt}{10pt}\selectfont\textbf{{\LARGE Cosmic ray interactions in the solar system:} \\ The Gerasimova-Zatsepin effect}} % Article title


\author{J.V.R. van Eijden, S.J. de Jong, C.J.W.P. Timmermans\\ \today}

\address{Nikhef and Radboud University Nijmegen, \\ P.O. Box 9010, 6500 GL, Nijmegen, The Netherlands}

\begin{abstract}
The Gerasimova-Zatsepin effect of collisions of ultra-high-energy cosmic ray nuclei with photons emitted by the sun may cause two simultaneous air showers on Earth. This effect is simulated using the full energy spectrum of solar photons, ray tracing through the interplanetary magnetic field and upper limit values for the iron and oxygen cosmic ray fluxes. Only the most abundant interactions in which a single proton is emitted from the nucleus are considered. For the first time the distributions of distances between the individual showers at Earth as a function of the distance of the primary cosmic ray to the Sun are shown. These distributions are used to estimate the capabilities of current detector arrays to measure the Gerasimova-Zatsepin effect and to show that a dedicated array is capable of measuring this effect.
\end{abstract}

\begin{keyword}
Cosmic rays \sep Spallation \sep Gerasimova-Zatsepin \sep Ray tracing \sep Distance distribution



\end{keyword}

\end{frontmatter}


\section{Introduction}
A heavy high energetic cosmic ray nucleus entering our solar system may interact with a photon from the sun, thereby emitting a particle.  
This is the so called Gerasimova-Zatsepin (GZ)  effect \cite{GZ}. 
When both secondaries from such an interaction hit the Earth, the effect is measurable. 
Different estimates of this effect have been presented in the past, e.g. \cite{MeTanco, Epele, Lafebre, Iyono, Anderson}.
Up until now no experimental evidence of the GZ effect has been reported. 
In this work, we describe a simulation of the GZ effect taking the full solar photon spectrum and interplanetary magnetic field into account. For the first time a ray tracing algorithm is used to generate accurate results  for the detectability of the GZ effect. From the results obtained by these simulations it is clear why current experiments have not been able to detect the GZ effect. We also use the measurable properties of this effect on Earth to design an experiment that is able to use the GZ effect to study the composition of the cosmic ray flux in space.

%------------------------------------------------

\section{GZ event simulation}

The GZ event simulation convolutes the energy spectrum of photons from the sun and the cosmic ray energy flux for nuclei with the cross section of photon-nuclei interactions to predict differential event rates. The fragments potentially hit the Earth are then back tracked using a ray tracing algorithm through the solar inter-planetary magnetic field to generate efficiently events for which both collision fragments hit the Earth. These elements will be discussed in the following subsections.

\paragraph{1. The sun's differential photon density function}\hfill \break
Planck's formula for black body radiation can be written as,
\begin{equation}
 {\rm d}n(\epsilon, T, r)= \left(\frac{r_{\odot}}{r}\right)^2\cdot \frac{2\pi}{c^3 h^3}\frac{\epsilon^2}{{\rm e}^{\epsilon/k_BT}-1} {\rm d} \epsilon ,
\end{equation}
with $n$ and $\epsilon$ the photon density and energy and $r$ the distance from the sun. All constants, speed of light,  $c$, Planck's constant, $h$, Boltzmann's constant, $k_B$ and radius of the sun, $r_{\odot}$, can be combined into a single scaling factor. 
At a distance $r$ of one astronomical~unit, using a temperature of $T=5771$~K for the photosphere of the sun, one obtains,
\begin{equation}
{\rm d} n(\epsilon,r)= 7.13 \cdot10^7\cdot \left(\frac{1}{r}\right)^2\cdot {\frac{\epsilon^2}{{\rm e}^{\epsilon/0.497}-1} {\rm d} \epsilon ,} \label{eq:dens}
\end{equation}
with ${\rm d} n$ the number of photons of energy $\epsilon$ (in eV) per cm$^3$ per eV and $r$ the distance from the sun in astronomical units.
The normalization constant is in agreement with \cite{MeTanco}, in which it was derived as a normalization factor to reproduce the correct photon flux as measured in experiments. \\
The photon density peaks at a photon energy of $\epsilon = 0.79$~eV. 
In our simulation we integrate over photon energies ranging from $0.04$ up to $24$~eV. Assuming that the sun is a perfect black body radiator this accounts for more than 99.9~\% of the solar spectrum.

\begin{figure}[H]
%%\begin{flushleft}
\includegraphics[width=0.5\textwidth]{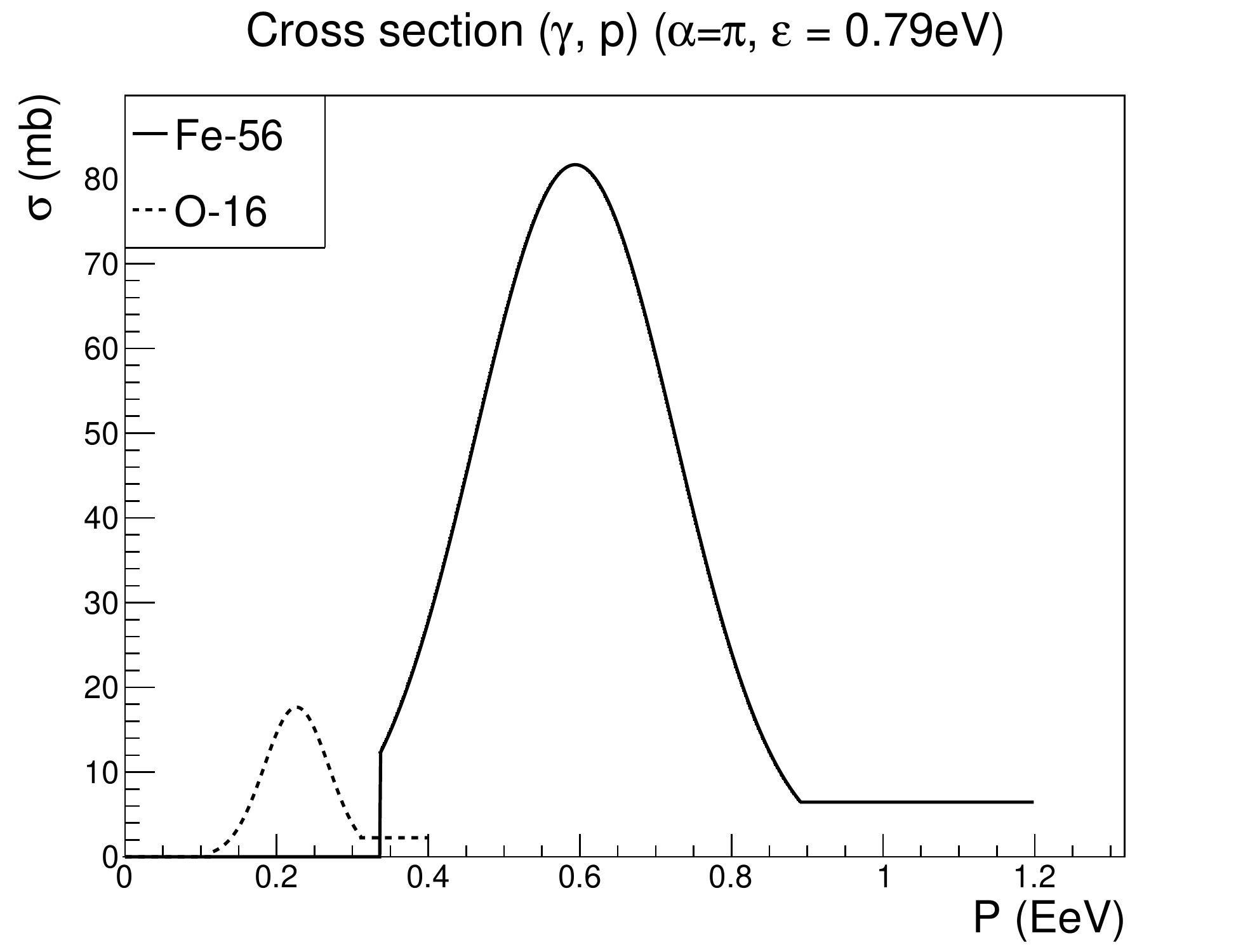}
\caption{Photon nucleus cross section of the Giant Dipole Resonances as a function of the nucleus energy in the rest frame of the sun. The full line to the right is for iron nuclei and the dotted line to the left is for oxygen nuclei. The photon energy is fixed at 0.79~eV and head-on collisions are assumed.}
\label{fig:Crosssections2}
%%\end{flushleft}
\end{figure}

\paragraph{2. Cross section} \hfill \break
We only consider the giant dipole resonance (GDR) as modelled in \cite{Stecker} and \cite{Puget}.
This is the dominant part of the total inelastic cross section just above threshold (see e.g.~\cite{GDPR}).
The used values are estimated to have an overall uncertainty of $\pm$~15\% based on total cross section measurements \cite{Puget}. As an example, \autoref{fig:Crosssections2} shows the partial cross sections for head-on collisions for iron and oxygen respectively with a 0.79~eV photon $\left( \gamma \right)$, in which a proton $\left(p \right)$ is expelled.  The cross sections are maximal at an energy of $\epsilon =0.59$~EeV and 
$\epsilon = 0.23$~EeV for iron and oxygen respectively. In the GZ event simulation the angle between the photon and incoming cosmic ray nucleus is determined at each point in space assuming the photons are emitted radially by the sun and the cosmic rays come from random directions. Furthermore, at each point we integrate over the photon energies.

\paragraph{3. Cosmic ray flux}\hfill \break
The cosmic ray flux is implemented by using a model of the differential energy spectrum \cite{Nagano},
\begin{equation}
J=C \cdot \left( \frac{E}{6.3 \cdot 10^{18}} \right)^{-3.20 \pm 0.05}
\end{equation}
for $4\cdot 10^{17} {\rm eV}<E<6.3\cdot10^{18} \rm{eV,}$\\
with\\
\begin{tabular}{l l}
J & differential flux $\rm{km}^{-2}\rm{s}^{-1}\rm{sr}^{-1}\rm{eV}^{-1}$,\\
C & $\left(9.23 \pm 0.65 \right) \cdot 10^{-27} \rm{km}^{-2}\rm{s}^{-1}\rm{sr}^{-1}\rm{eV}^{-1}$, and\\
E & the energy of the cosmic ray nucleus in eV.\\
\end{tabular}\\
The composition in the studied energy range is dominated by heavy nuclei \cite{Kampert}. As typical elements we investigate oxygen and iron.

\begin{figure}[H]
%\begin{flushleft}
\includegraphics[width=0.5\textwidth]{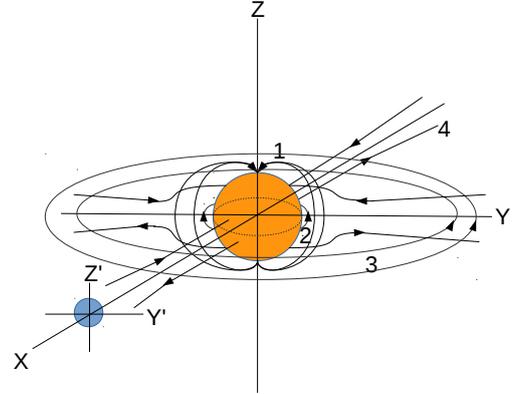}
\caption{Solar magnetic field consisting of four main components 1. Dipole field, 2. Sunspots field, 3. Dynamo field, and 4. Ring field.}
\label{fig:Magfield}
%\end{flushleft}
\end{figure}

\paragraph{4. Magnetic fields}\hfill \break
The magnetic field models described in \cite{Akasofu} and \cite{Epele} are used in our simulation. In these models, the total field consists of four components, 
as sketched in \autoref{fig:Magfield}: the dipole, sunspots, dynamo and ring fields.\\
The total field, responsible for the deflection of charged particles, is obtained by adding the different components. The strength of the modelled magnetic field at one Astronomical unit is $B_{total}=5.3 \cdot 10^{-5}$~G. This is at the high end of the magnetic field strength, at Earth, during `normal' solar activity which ranges from about 2 to 5~nT \cite{Wang}.\\
To save CPU-time a ray tracing algorithm was used to calculate the tracks of particles that can actually hit the Earth. First the path is calculated of the particles starting at Earth with an outward momentum and inverse charge. At a sphere of 4~AU around the sun the direction and charge are reversed and interaction probabilities are taken into account when tracking from this 4~AU sphere back to Earth.

%------------------------------------------------

\section{Detector simulation}
We focus on two different detector configurations: the Pierre Auger Observatory \cite{Auger} as the largest, $50\times 60$~km$^2$, ultra-high-energy cosmic ray observatory, but with a relatively large energy threshold and the HiSPARC distributed detector that also covers large distances and has a lower energy threshold, but a much smaller effective area. Since the events and detector are simulated independently any other detector configuration can easily be implemented in our scheme.

\begin{figure}[H]
%\begin{flushleft}
\includegraphics[width=0.5\textwidth]{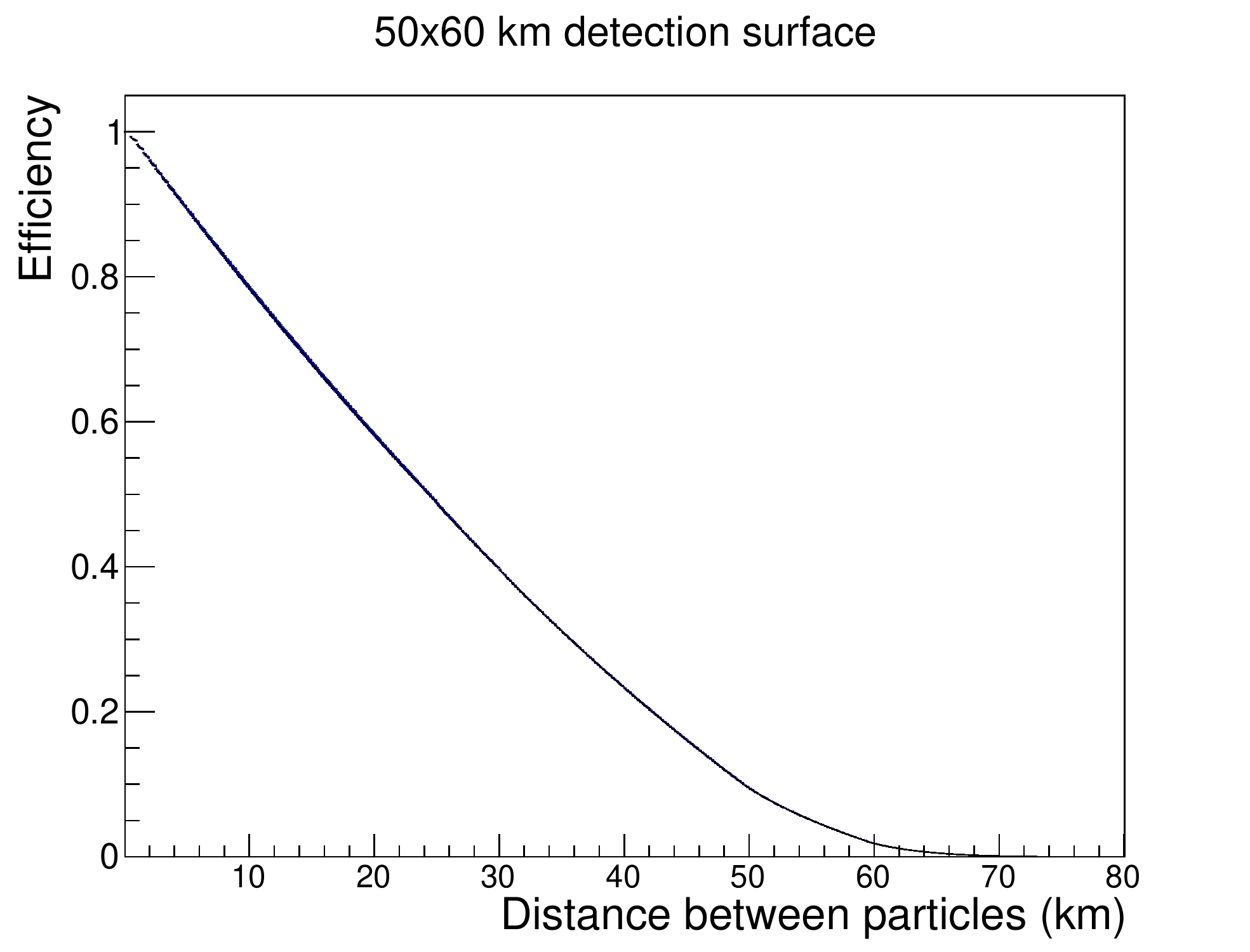}
\caption{The probability for the second secondary to fall into a $60 \times 50$~km area once the first secondary is detected anywhere in the same particle area.}\label{fig:Fraqdetec}
%\end{flushleft}
\end{figure}

\paragraph{1. The Pierre Auger Observatory} \hfill \break
The Pierre Auger Observatory is modeled as $50 {\rm x} 60~{\rm km}^2$ covered with $1.5$~km spaced $10$~m$^2$ particle detectors. 
Given that both secondaries hit the Earth, the probability that the first secondary hits this surface is given by the ratio of the detector surface over the surface of the Earth. The probability that the second secondary hits this surface depends on distribution of the separation distances between the two particles. 
The probability that the second secondary will hit the detector is calculated as a function of the separation distance between the two particles, see \autoref{fig:Fraqdetec}. The detection rate is obtained by multiplying this distribution with the distribution of the distance between the particles on Earth.

\begin{figure}[H]
%\begin{flushleft}
\includegraphics[width=0.5\textwidth]{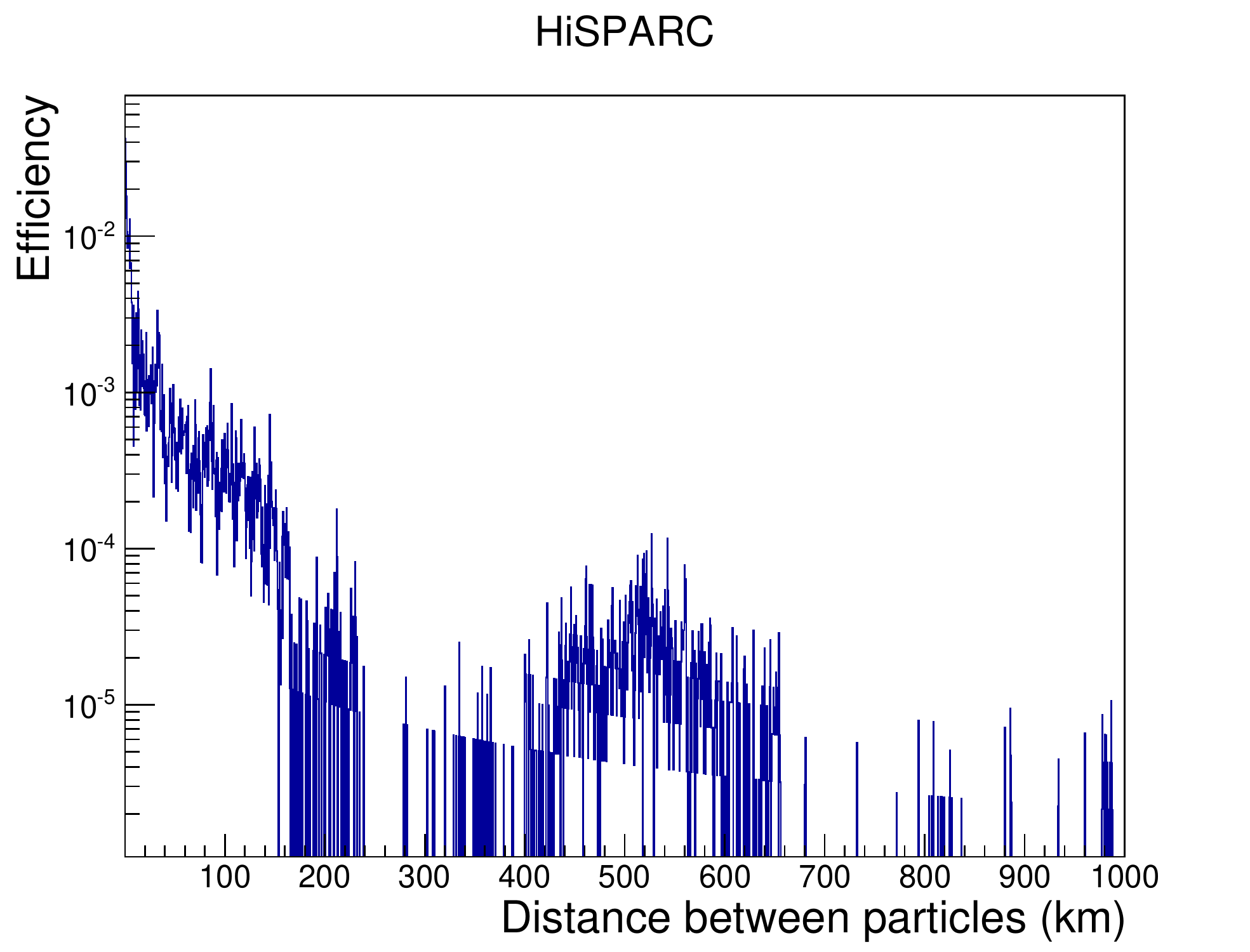}
\caption{Probability to detect the second secondary with HiSPARC configuration given that the first secondary has been detected by HiSPARC.}\label{fig:HiSPARC2}
%\end{flushleft}
\end{figure}

\paragraph{2. HiSPARC} \hfill \break
HiSPARC stands for High School Project on Astrophysics Research with Cosmics. The detector array consists currently of 119 detector stations placed on participating high schools and universities. 
In our calculation we assume that a station will be triggered by showers up to distances of $0.5$~km. In \cite{Fokkema} it is shown that this is a reasonable assumption, even though in reality, this distance will depend on the energy of the shower and the cutoff is soft. Using the single station detection area, the total effective detection surface of HiSPARC is found to be  $93~\rm{km}^2$. The probability that the first secondary hits HiSPARC is given by the ratio of the total detection surface over the surface of the Earth. The probability that also the second secondary hits HiSPARC depends on the separation distance between the two particles. To calculate this probability the surface distribution of all combinations of stations is used, as shown in  \autoref{fig:HiSPARC2}.

%------------------------------------------------

\section{Results}
\begin{figure}[H]
%\begin{flushleft}
\includegraphics[width=0.45\textwidth]{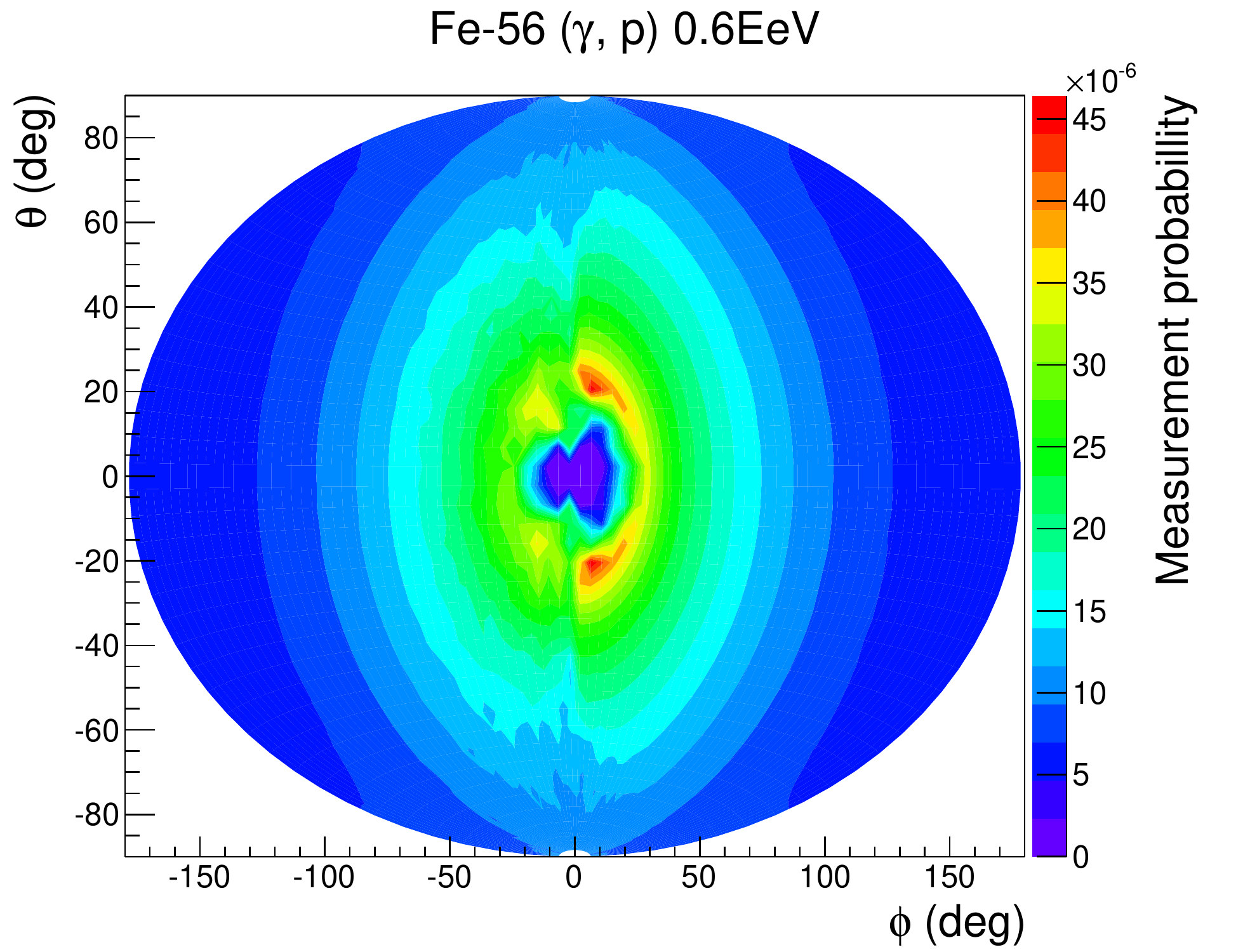}
\caption{Probability of observing a GZ event from a interaction of an iron nucleus of 0.6~EeV on Earth as a function of the arrival direction of the heaviest secondary. The sun is located at (0,0).}\label{fig:PMFe06EeV}
%\end{flushleft}
\end{figure}

\begin{figure}[H]
%\begin{flushleft}
\includegraphics[width=0.45\textwidth]{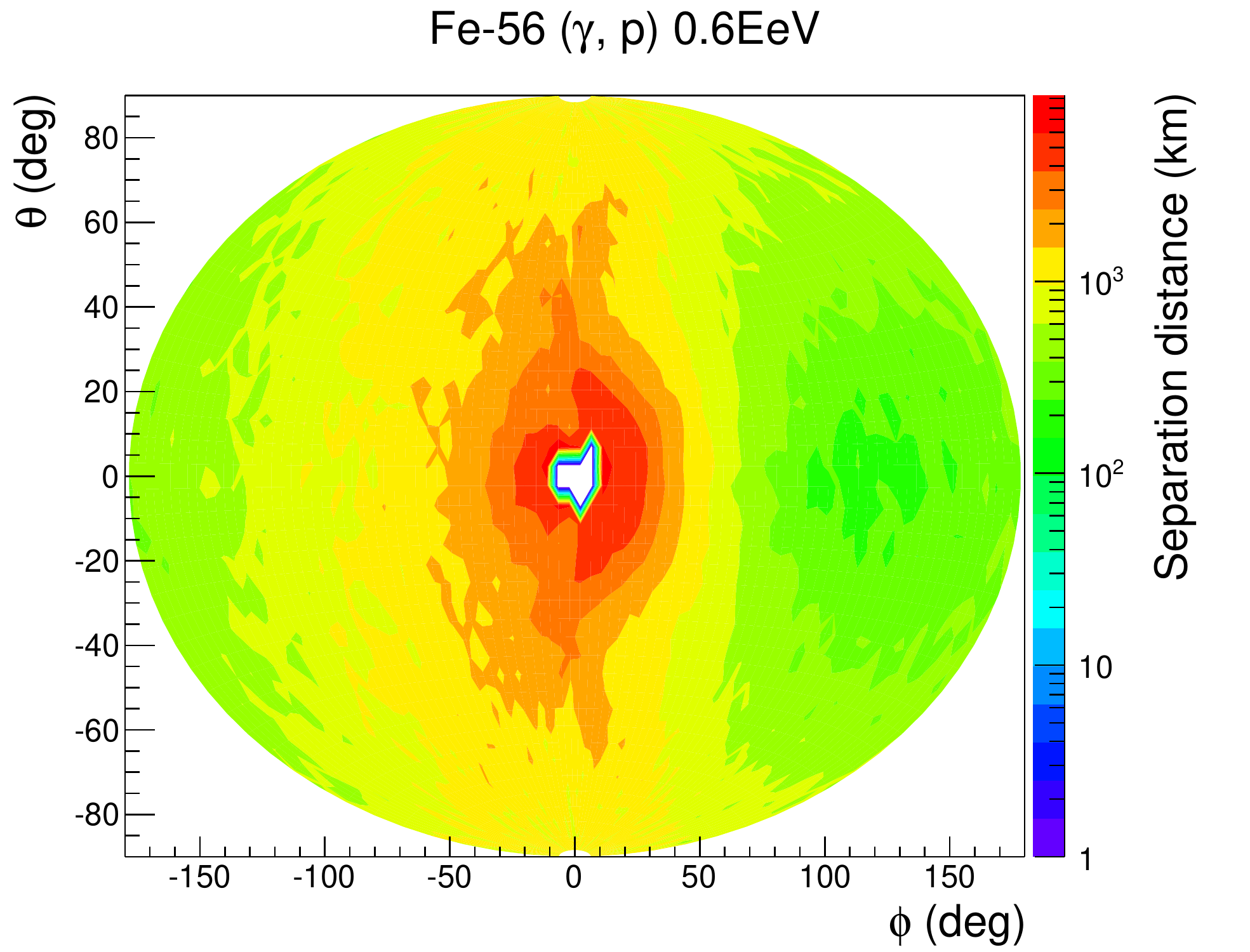}
\caption{Average separation distance in km of the secondaries from a GZ interaction of an iron nucleus of 0.6~EeV on Earth as a function of the arrival direction of the heaviest secondary. The sun is located at (0,0).}\label{fig:AsepFe06EeV}
%\end{flushleft}
\end{figure}

Using our simulation, we have studied the probability of measuring the GZ effect as a function of the arrival direction of the heaviest secondary on Earth.  As an example,  \autoref{fig:PMFe06EeV}  shows the result for iron nuclei of an energy of $0.59$~EeV. In this Figure, the sun is located in the center and $\phi$ describes the angle in the Earth-Sun plane. The angle perpendicular to the plane is denoted as $\theta$.\\
In order to identify the GZ effect, both resulting particles have to be measured. Therefore, the detector needs to be large enough for such an observation.
\autoref{fig:AsepFe06EeV} shows the average distance between both secondaries as a function of the arrival direction of the heaviest secondary on Earth. Again the sun is located in the center. Combining Figures \ref{fig:PMFe06EeV} and \ref{fig:AsepFe06EeV}, it is clear that the probability for interacting is largest when the nucleus passes the sun at a small distance. However, even if both fragments resulting from this interaction arrive at Earth, the distance between them is large.\\

\begin{figure}[H]
%\begin{flushleft}
\includegraphics[width=0.5\textwidth]{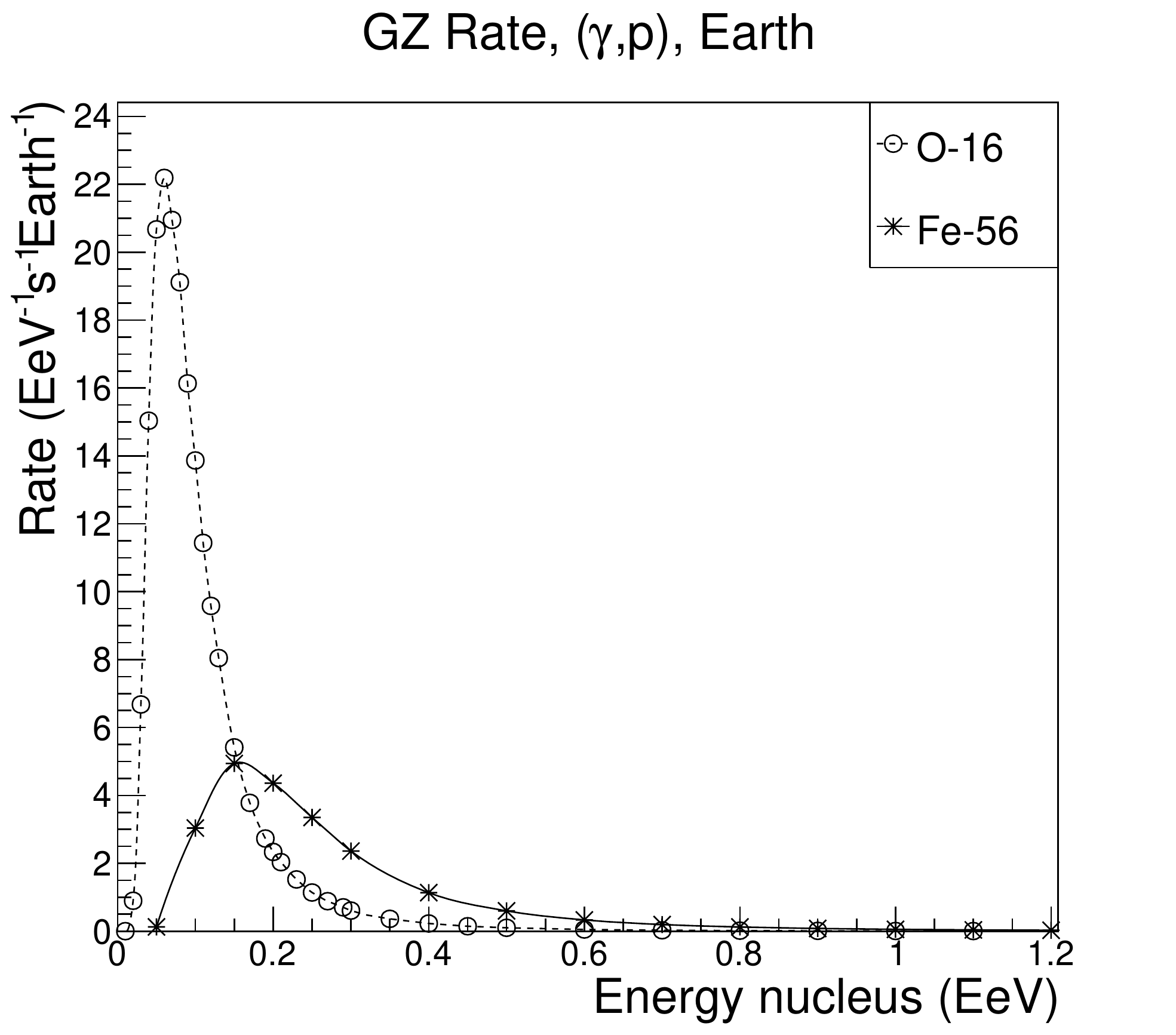}
\caption{Differential rate of GZ events reaching the Earth as a function of energy, assuming that the full flux consists of oxygen and iron respectively. The connecting lines are only plotted to guide the eye.}\label{fig:FeOx}
%\end{flushleft}
\end{figure}

\noindent \autoref{fig:FeOx} shows the results of a full  simulation of the GZ rate at Earth for iron and oxygen primaries as a function of incoming energy. In this simulation the complete cosmic ray flux is assumed to consist of either iron or oxygen primaries, therefore the values can be treated as upper limits. The energy at which the differential rate is maximal is around $0.15$~EeV and $0.06$~EeV, for iron and oxygen respectively. These energies are below the maxima shown in  \autoref{fig:Crosssections2} due to the steepness of the cosmic ray spectrum.

\begin{figure}[H]
%\begin{flushleft}
\includegraphics[width=0.5\textwidth]{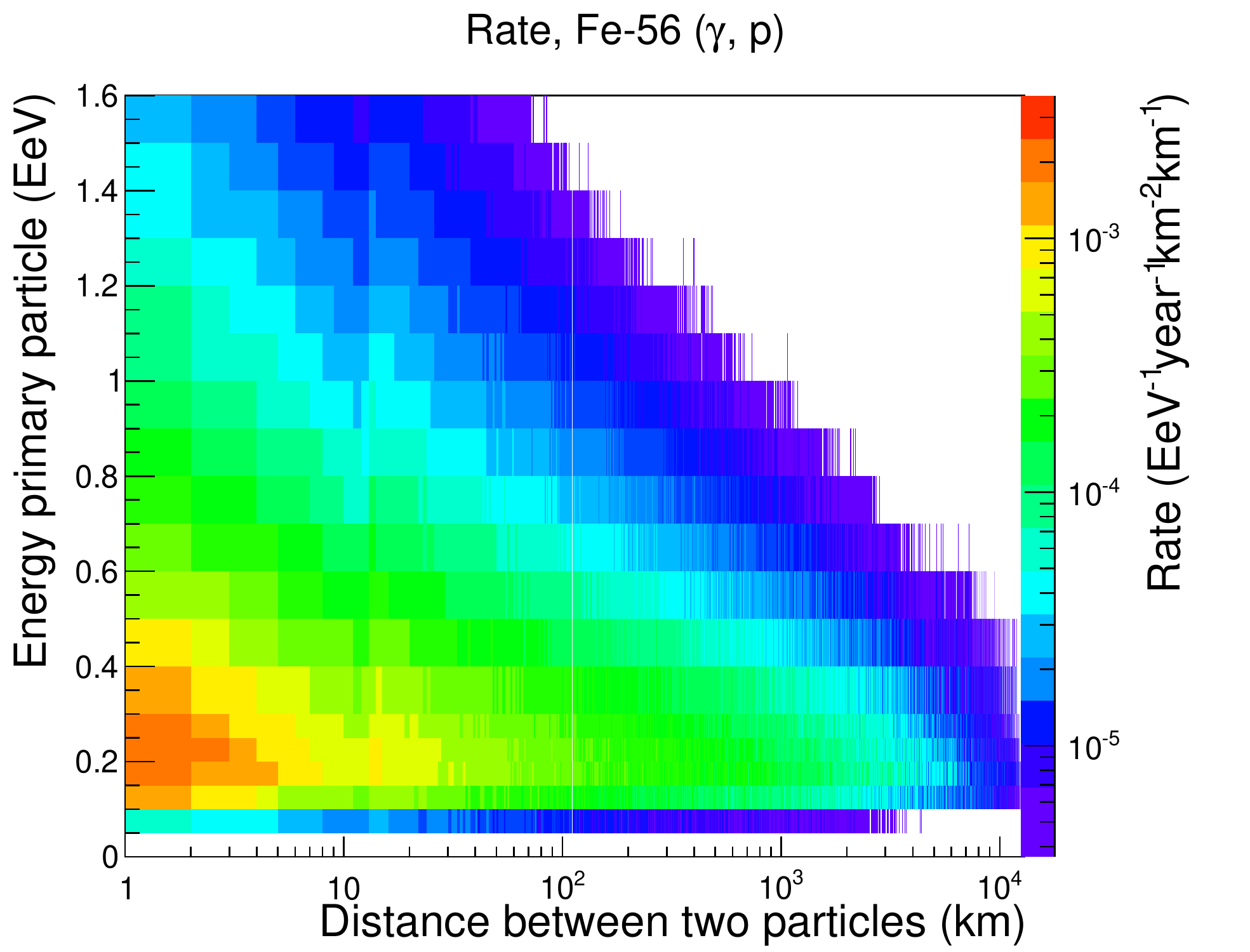}
\caption{Differential event rate as function of energy and separation distance if the flux entirely consists of iron nuclei.}\label{fig:FeDiff}
%\end{flushleft}
\end{figure}

\begin{figure}[H]
%\begin{flushleft}
\includegraphics[width=0.5\textwidth]{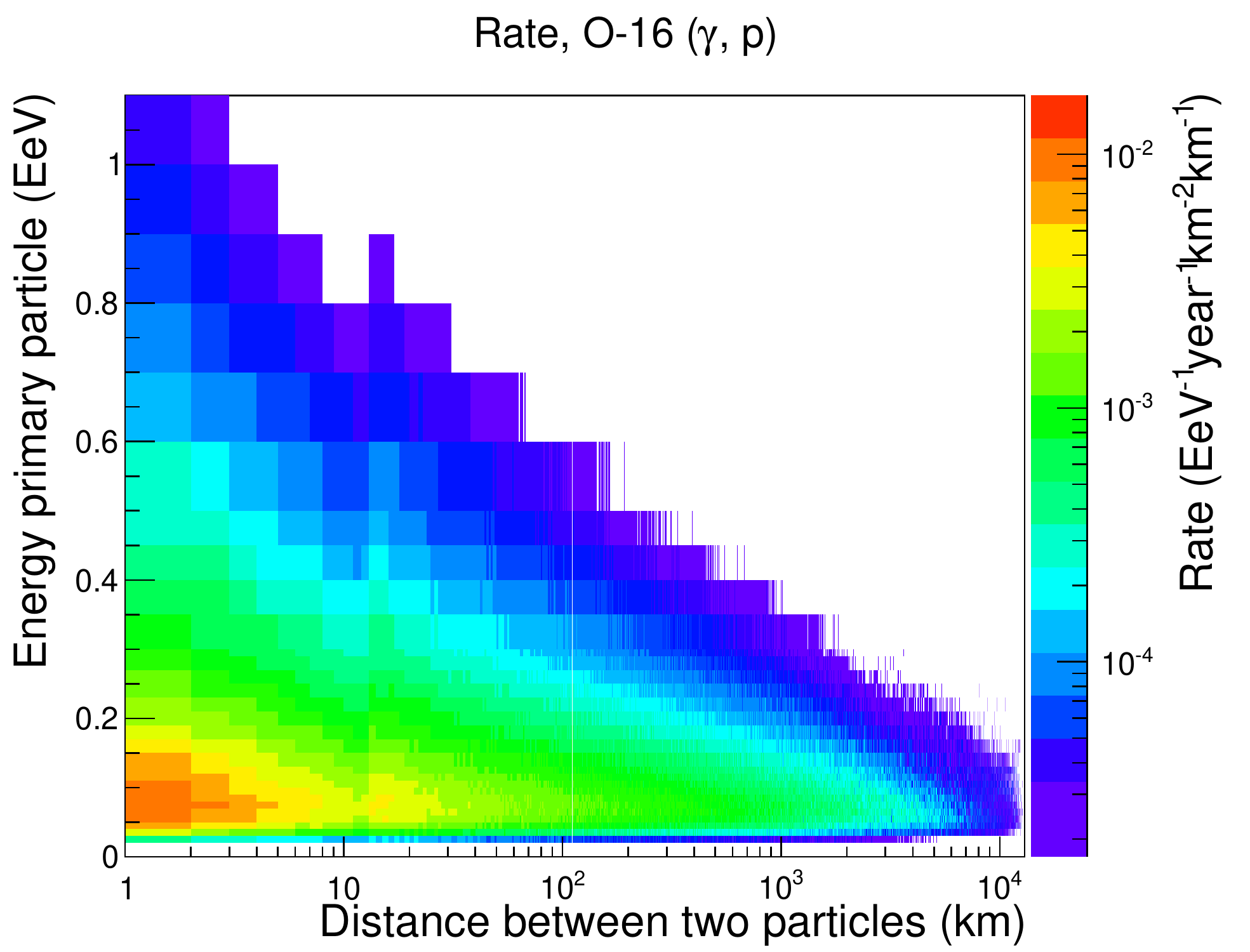}
\caption{Differential event rate as function of energy and separation distance if the flux entirely consists of oxygen nuclei.}\label{fig:OxDiff}
%\end{flushleft}
\end{figure}

\noindent The integration of the differential rate provides an upper limit of $0.87$ ($1.7$) GZ pairs hitting the Earth each second, assuming that the complete flux 
of cosmic rays consists of iron (oxygen).\\
\autoref{fig:FeDiff} and \autoref{fig:OxDiff} show the rate per square kilometer as a function of the distance between the two secondaries and the energy of the primary iron or oxygen nucleus.

\begin{figure}[H]
%\begin{flushleft}
\includegraphics[width=0.5\textwidth]{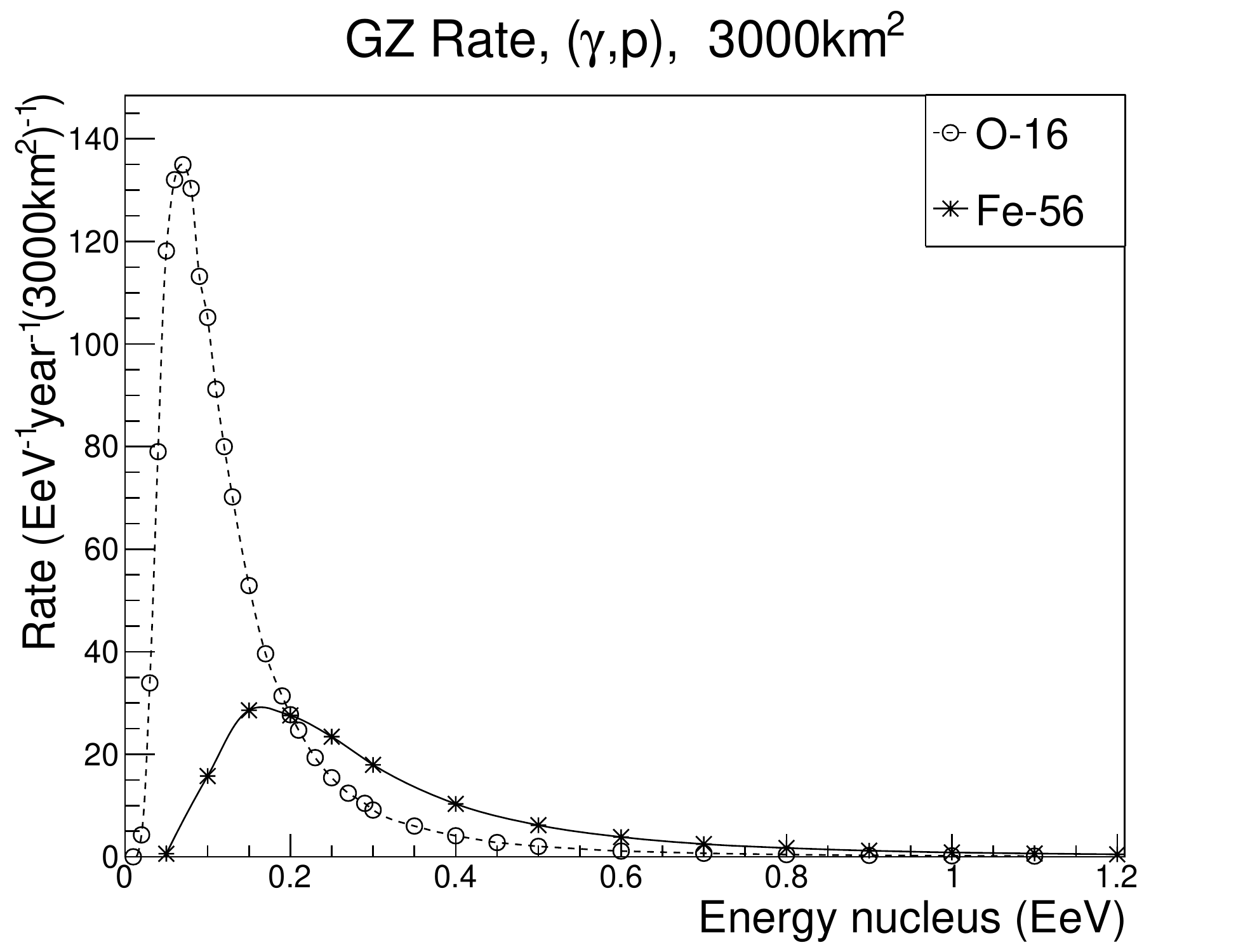}
\caption{Differential rate of GZ events recorded by a 50x60~km$^2$ detector
as a function of energy, assuming that the full flux consists of oxygen and iron respectively.}\label{fig:AugerFeOx}
%\end{flushleft}
\end{figure}

\begin{figure}[H]
%\begin{flushleft}
\includegraphics[width=0.5\textwidth]{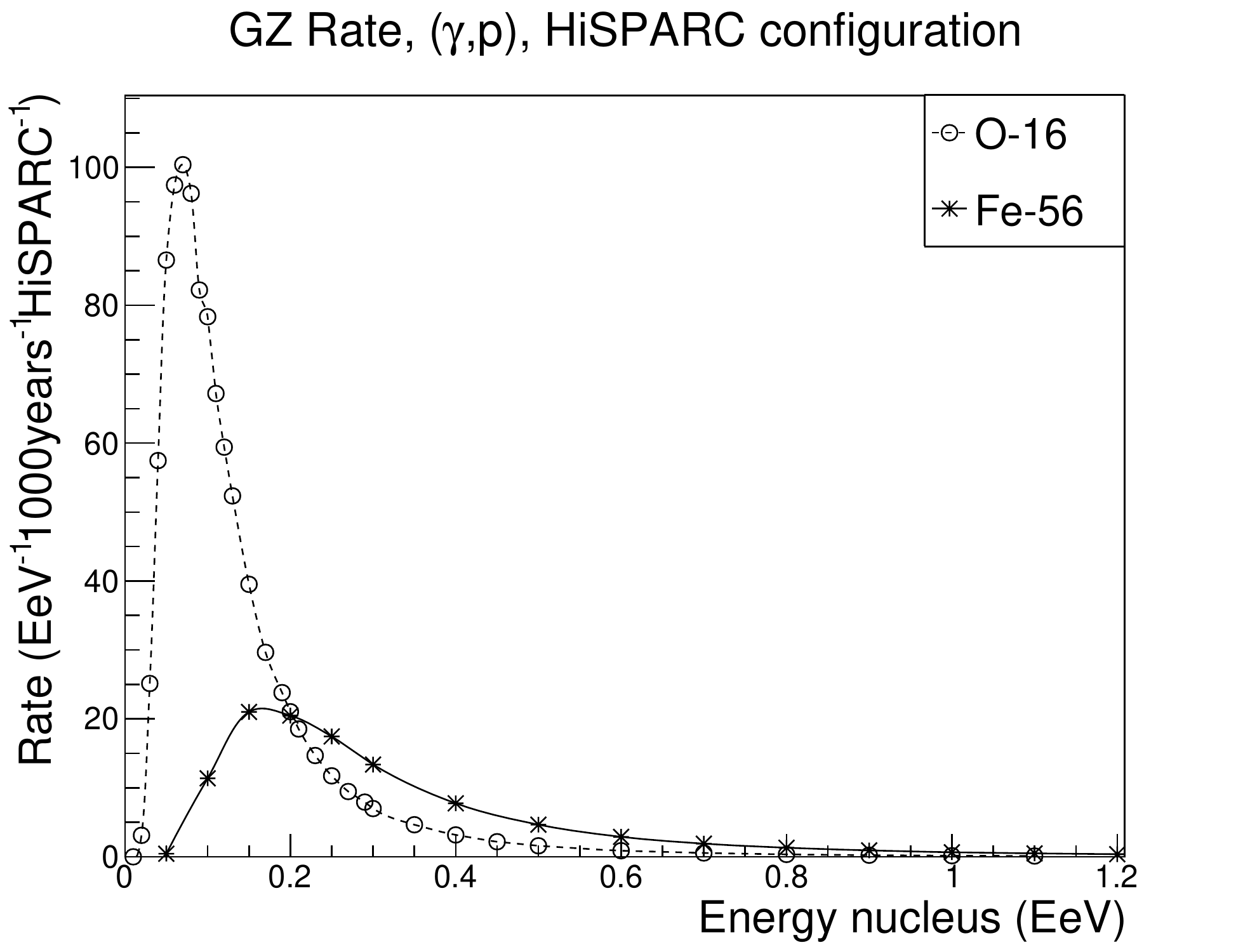}
\caption{Differential rate of GZ events for which the two secondaries de-
tected by HiSPARC, assuming that the full flux consists of oxygen and iron respectively.}\label{fig:HiSPARCFeOx}
%\end{flushleft}
\end{figure}

\noindent The maximal density of GZ events is found for small separation distances between the nuclei. Identification of individual nuclei at distances below~2\,km 
would be beneficial for this study. In general, for larger primary energies, the distance between the nuclei at Earth becomes smaller, which agrees well with our naive expectations.\\
Finally the results of a full simulation of the GZ rate at Earth, \autoref{fig:FeOx}, is used in combination with a detector simulation. \autoref{fig:AugerFeOx} shows the differential rate of GZ events per year for a 50x60~km$^2$ detector as a function of energy of the primary particle. \autoref{fig:HiSPARCFeOx} shows the differential rate of GZ events per 1000 years for a  HiSPARC configuration as a function of energy of the primary particle.

\noindent Integrating   \autoref{fig:AugerFeOx} provides an upper limit of $6.5$ (14) GZ pairs that hit a 3000\,km$^2$ detector every year, assuming
that the flux of cosmic rays consists completely out of iron (oxygen). The HiSPARC rates are three orders of magnitude lower, at 4.9 and 10 events every 1000
years for iron and oxygen respectively.

%------------------------------------------------

\section{Conclusions}

We have shown that the probability of a GZ interaction for cosmic nuclei passing through our solar system is maximally $4.5\cdot10^{-5}$. This maximum occurs when its trajectory is close to the surface of the Sun. However, the solar magnetic fields will cause a large deflection which makes it unlikely that both fragments are measurable at Earth. In this case it is much more favourable to perform a measurement of the GZ effect in a direction away from the Sun. The lower interaction probability is partly compensated by a lower energy threshold for the incoming cosmic rays, as the interaction with solar photons is head-on. Furthermore, the smaller solar magnetic fields cause a separation of both fragments on Earth that may be as small as a few kilometers, as shown in \autoref{fig:FeDiff} and \autoref{fig:OxDiff}. Our simulations show that the detection area of the HiSPARC experiment is too small to measure the GZ effect. Furthermore, we have shown that the energy threshold of the Pierre Auger Observatory is too high for such an observation \cite{Auger}. However, a  $3000$\,km$^2$ detector that is sensitive to both fragments will detect about 10~GZ~events each year. The detector will have to be fully efficient for showers with energies ranging between $10^{15}$\,eV and $10^{17}$~eV. The energies of the fragments provide an estimate of the energy of the original cosmic ray. In addition to the energy ratio of the fragments, the distance distributions given in \autoref{fig:FeDiff} and \autoref{fig:OxDiff} can be used to estimate the mass of the primary cosmic ray.

%------------------------------------------------
%------------------------------------------------

%%\section{}
%%\label{}





%%\break

\end{document}